# Intent Signal Theory: A Computational Framework for Intent-State Control in Human–AI Interaction

**Gang Peng**
Huizhou Lateni AI Technology Co., Ltd., Huizhou, China
Huizhou University, Huizhou, China
ORCID: 0009-0007-4774-1681
Correspondence: peng@hzu.edu.cn

## Abstract

Human–AI interaction is commonly modelled as a mapping from prompt text to model output. We argue that this framing omits a critical object: the user's source intent — the latent goal state that precedes and motivates the prompt. When source intent is treated as the object to be preserved, it becomes possible to ask not whether a response resembles the prompt, but whether it preserves what the user meant. Here we introduce Intent Signal Theory (IST), a computational framework that provides this missing intent layer. IST distinguishes four objects routinely conflated in practice: latent source intent $I*$, an observable intent proxy $\tilde{I}$, an encoded carrier P, and model output O. It formalises task-conditioned dimensional weights, encoding masks, structural and fidelity recovery scores, and intent drift. IST further decomposes intent dimensions into a public component recoverable from model priors and a private component requiring explicit encoding. The Theorem of Irreversible Intent Loss (TIIL) formalises that, under a fixed carrier-only information state, private intent absent from the carrier cannot be recovered beyond generic substitution. Grounding evidence from four companion empirical studies — spanning six LLMs, three languages and three task domains — shows structural-fidelity splits, human-validated metric dissociation, and weight-tolerance plateaus consistent with IST's predictions. IST reframes prompt engineering as intent-protocol design and identifies an explicit computational layer that current AI systems lack.

## 1. Introduction

The dominant paradigm in large language model (LLM) research treats interaction as a mapping from prompt text to generated output [1,2,3,4]. This framing has driven advances in instruction tuning, chain-of-thought elicitation, and RLHF-based alignment [5,6], all of which manipulate the surface form of the prompt or the training signal. Yet it rests on an implicit assumption rarely examined: that the prompt is the primary object of the exchange. We argue it is not.

Before any prompt is composed, the user holds a latent goal state — a source intent that preceded and motivated the act of prompting. The prompt is the carrier of this intent, not the intent itself. A model response is genuinely satisfactory not when it resembles the prompt surface but when it preserves the source intent the prompt was meant to encode. Consider a user who writes: "Write a competitive analysis for our product." The model produces a polished, well-structured analysis — comprehensive, coherent, preferred by holistic evaluators — directed at the consumer market. But the user's unstated intent specified enterprise clients, a different competitive landscape, and proprietary benchmarking criteria. The output is structurally excellent and intent-deficient. No surface-level prompt critique identifies the failure; no holistic quality score detects it. The problem lies not solely in the model, but in the gap between what was encoded in the carrier and what the user meant.

This gap — between prompt surface and source intent — is systematic, not incidental. It arises because users cannot always fully verbalise their goal states [13,14,15,22], because they assume models will infer what is left implicit [23], and because current evaluation instruments measure output quality against the prompt surface rather than against the latent goal. When holistic evaluation assigns high scores to outputs that fill expected structural slots while substituting generic defaults for the user's private, individual-specific content, it systematically obscures the failure mode that matters most.

Prompt engineering treats effective expression as an empirical target rather than a theoretical object [4,30,31]. Alignment research measures conformity to holistic human preference [5,6], which cannot distinguish structural coherence from dimensional intent fidelity. Hallucination research addresses epistemic failures on the model side [7,8,9] — content unsupported by training data or retrieved context — but leaves encoding failures on the user side unaddressed. Evaluation frameworks aggregate output quality into scalar scores [10,11], conflating structural completeness with intent reproduction. Across these programmes, a common absence emerges: there is no unified computational account of what source intent is, how it enters the prompt, which parts the model can recover from its prior, and which parts are irreversibly lost once omitted from the carrier.

This is the missing intent layer in human–AI interaction. Intent Signal Theory (IST) is a theory of that layer.

IST models single-turn human–AI interaction as intent-state transmission and control: a process in which a latent source intent is approximated by an observable proxy, encoded into a carrier, partially recovered by the model, and evaluated for fidelity. The framework introduces a set of formal objects — dimensional weights, encoding masks, structural and fidelity recovery scores, public–private decomposition — that make intent-state control computationally tractable. It is grounded in four companion empirical studies, all publicly available on arXiv, that form a structured evidence chain [12,19,20,21]. IST provides a unified theoretical account of why structured intent encoding improves alignment, why the effect generalises across languages and models, why structured intent behaves as a protocol-like mechanism, and why holistic evaluation instruments systematically fail to detect intent fidelity loss. Its scope is explicitly bounded to single-turn LLM generation; multi-turn dialogue, agentic tool use, and multimodal extensions are prospective directions that require separate formalisation.

## 2. A Computational Framework for Intent-State Control

By *intent-state control* we mean the explicit management of how source intent is represented, encoded, recovered, measured and audited within a single generative episode. The framework introduced here makes each of these operations computationally tractable.

### 2.1 Four objects and the transmission chain

IST is built on the recognition that four objects routinely conflated in LLM research are theoretically distinct. The **latent source intent** $I*$ is the goal state the user holds at the outset of an interaction — the fixed reference against which encoding loss and output fidelity are defined. It is not directly observable; empirical work approximates it through observable intent proxies. The **observable intent proxy** $\tilde{I}$ is a structured, operationalisable approximation of $I*$ that can be decomposed into dimensions and scored against model output. The **encoded carrier** P is the actual prompt that enters the model's processing chain — the only intent-bearing signal explicitly supplied for the current generation. The **model output** O is generated from P through learned parameters and inference. The transmission chain

$$I* \rightarrow \tilde{I} \rightarrow P \rightarrow O$$

is the theoretical backbone of IST. This formulation is distinct from Shannon's information theory [16,17]. Whereas Shannon's framework formalises signal transmission in communication systems, IST addresses the representation, encoding, recovery and evaluation of user-specific intent states in human–AI interaction. Each link is a site of potential information loss, weight mismatch, or partial compensation from the model prior. Figure 1 summarises the resulting transmission chain and the missing intent layer introduced by IST.

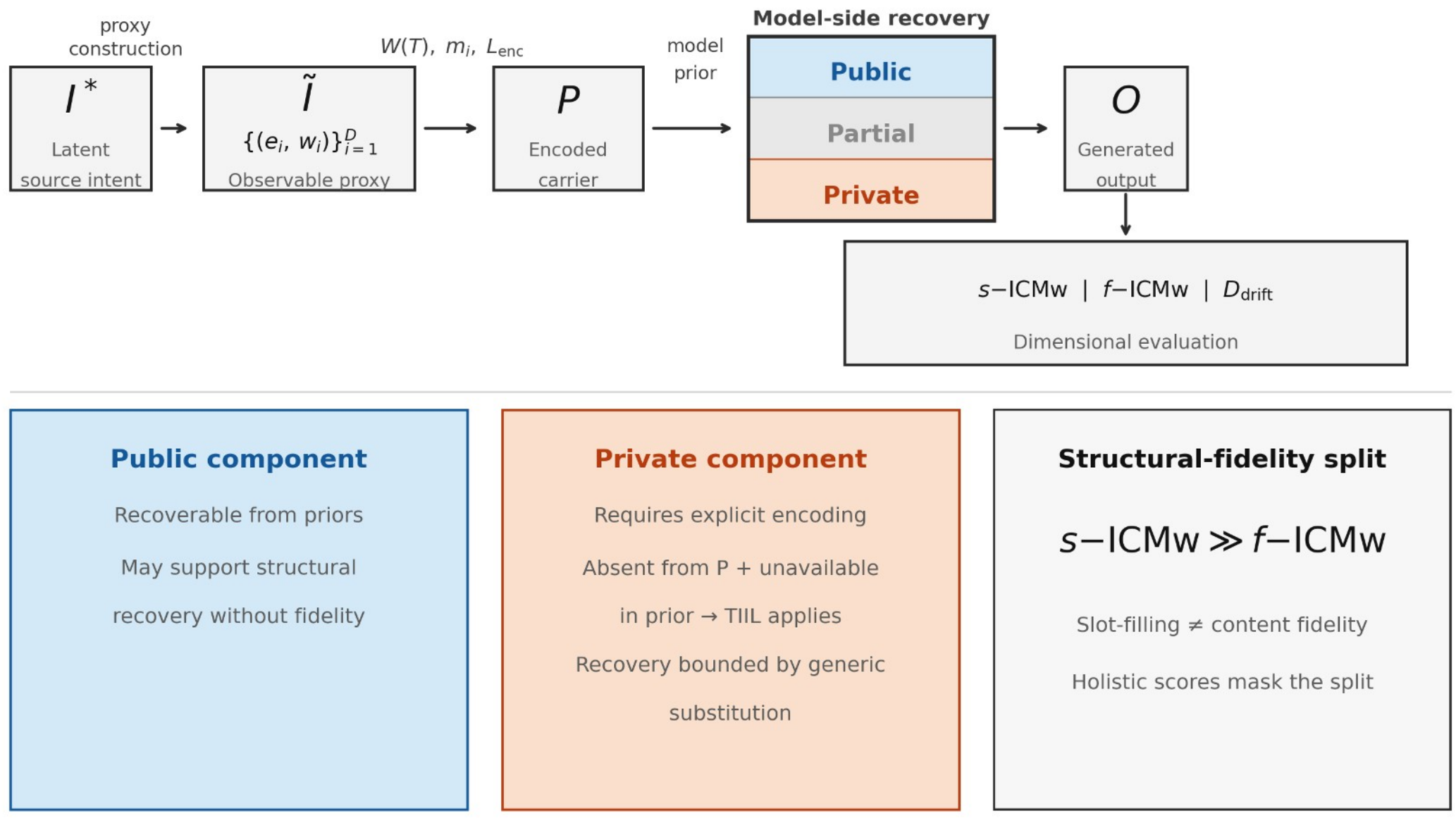

**Figure 1 |** The missing intent layer in human–AI interaction. IST positions an explicit computational intent layer between the user's latent source intent $I*$ and the model output O. The chain $I* \rightarrow \tilde{I} \rightarrow P \rightarrow O$ distinguishes source intent, observable intent proxy, encoded carrier and generated output. Model-side recovery may restore public or partial structure from priors, whereas private components require explicit encoding. Dimensional evaluation separates structural recovery from fidelity recovery, revealing structural–fidelity splits that holistic evaluation can conceal.

## 2.2 Foundational assumptions

IST is built on two structural assumptions whose adequacy can be assessed empirically. **F1 (Source intent integrity)**: at the outset of a single-turn interaction, the user holds a source intent $I*$ that serves as the fixed reference object for the interaction — a goal state that exists prior to encoding, with respect to which the output can in principle be evaluated. **F2 (Dimensional decomposability)**: for modelling and evaluation, the observable proxy $\tilde{I}$ can be represented as a finite weighted set of dimensions whose composition is a modelling choice, not a property of a unique natural decomposition. Two further modelling conventions are adopted: hierarchical refinability (any dimension may be decomposed into sub-dimensions without altering the formalism) and encoding independence (structuring intent increases the explicit carryability of existing intent without creating new intent). These assumptions are not axioms claimed to be self-evidently true; they are modelling commitments whose empirical adequacy is evaluated by whether IST's predictions are borne out.

## 2.3 Dimensional decomposition and task-conditioned weights

For a given task type T, the observable proxy is represented as a finite weighted set of dimensions:

$$\tilde{I} = \{(e_i, w_i)\}, \quad w_i \geq 0, \quad \Sigma w_i = 1$$

The weights $w_i$ are task-conditioned: different tasks assign different importance to different intent dimensions, and the fidelity consequences of omitting a dimension depend on its weight under the task, not on any dimension-intrinsic property. This normalization is a modelling convention, not a claim that latent source intent has a natural unit or measurable magnitude; it establishes a shared scale across conditions, models, and task domains. In practice, weight distributions can be estimated through systematic ablations — removing one dimension at a time and measuring the fidelity impact on model output [12,19]. The 5W3H/PPS framework [19,20,21] — decomposing intent across eight named dimensions (What, Why, Who, When, Where, How-to-do, How-much, How-feel) — is one high-coverage instantiation used as an operational testbed; IST does not require it and applies to any decomposition scheme that satisfies F2.

### 2.4 Encoding masks and encoding loss

An encoding mask $m_i \in \{0,1\}$ indicates whether dimension $e_i$ is explicitly represented in the carrier ($m_i = 1$) or absent ($m_i = 0$). Observable encoding loss is:

$$L_enc = 1 - \Sigma_i w_i m_i$$

This quantity measures the share of task-weighted intent not explicitly encoded in the prompt. A value of zero means the carrier encodes all task-relevant dimensions at full weight; positive values indicate that some task-weighted intent is absent and must be recovered — if at all — from the model's prior.

### 2.5 Structural recovery, fidelity recovery, and intent drift

IST distinguishes two recovery scores. The **structural recovery score** $r_i \in [0,1]$ measures the degree to which the output fills the expected dimensional slot, irrespective of whether the content is the user's intended value. The **fidelity recovery score** $f_i \in [0,1]$ measures the degree to which the output reproduces the user's specific intended content. Weighted aggregates:

$$\text{s-ICMw} = \Sigma_i w_i r_i, \qquad \text{f-ICMw} = \Sigma_i w_i f_i$$

The central claim of IST is that these quantities can diverge: s-ICMw ≫ f-ICMw is the signature of the structural-fidelity split — the model restores the structural scaffold of the intent while replacing private content with high-frequency defaults from its training distribution (Figure 2 illustrates this split with a concrete example). Observable intent drift is:

$$D_drift = 1 - \text{f-ICMw}$$

### 2.6 Public–private decomposition

When a dimension $e_i$ is absent from the carrier, whether recovery succeeds depends on a property of the model prior: whether the intended value is predictable from aggregate training patterns. We partition intent dimensions relative to a given task context and model prior.

The **public component** consists of dimensions whose intended values are sufficiently predictable from the model's training distribution to produce contextually appropriate output. The **private component** consists of dimensions whose intended values are specific to the individual user — personal preferences, idiosyncratic constraints, or contextual requirements that deviate from aggregate defaults. For these dimensions, the model may supply a plausible generic default that fills the structural slot while failing to reproduce the user's intended value: this is the mechanism generating the structural-fidelity split.

The classification is operational, not ontological: a dimension is public or private relative to a task type, model prior, and specification context. Empirical estimation proceeds through ablation — removing each dimension in turn and measuring fidelity recovery against a protocol-complete observable proxy.

## 3. The Theorem of Irreversible Intent Loss

**Box 1 | Irreversible private-intent loss under carrier-only information**

*Assumptions.* (A1) Single-turn interaction: the model has access only to the encoded carrier P and its trained prior M, with no external memory, retrieval, or prior turns. (A2) Absent encoding: dimension $e_k$ is absent from the carrier, $m_k = 0$. (A3) Private classification: the intended value $v_k$ of $e_k$ is not recoverable from M at the specificity required for fidelity — formally, $I(v_k; P_{-k}, M) \approx 0$ at required specificity, where $P_{-k}$ denotes the carrier with $e_k$ absent.

*Statement.* No decoder g operating solely on $P_{-k}$ and M can recover $v_k$ beyond generic substitution under this fixed information state.

*Information-theoretic rationale.* Under the Markov structure of the IST transmission chain — $v_k \rightarrow (P_{-k}, M) \rightarrow g(P_{-k}, M)$ — any decoder operating in the fixed carrier-only information state has access only to $(P_{-k}, M)$. The data processing inequality [18,24,25] gives, for any decoder g:

$$I(v_k;\ g(P_{-k},\ M)) \ \leq\ I(v_k;\ P_{-k},\ M)$$

By assumption A3, $I(v_k; P_{-k}, M) \approx 0$ at the required specificity. Therefore $I(v_k; g(P_{-k}, M)) \approx 0$: the generated output cannot contain information about $v_k$ beyond the generic-substitution baseline. The "≈ 0" denotes mutual information insufficient to identify $v_k$ above chance at the required specificity; it is not absolute zero. The inequality applies to any decoder architecture — it is a property of the information available in the input state, not of the decoder's capacity.

*Boundary conditions.* TIIL holds when (A1)–(A3) are jointly satisfied. Multi-turn settings, retrieved context, user memory profiles, or a richer prior that reclassifies $e_k$ as public all lie outside the theorem's scope. TIIL applies to the specific intended value of $e_k$, not to the structural slot: a model may fill the slot with a plausible generic value while failing to recover $v_k$.

*Practical consequence.* Model scaling does not eliminate the private intent gap. A more capable model M' may expand the public component — making previously private content recoverable — but cannot recover $v_k$ if $v_k$ remains absent from the carrier and unavailable in the effective prior. The private boundary migrates as models scale, but it does not vanish for genuinely individual-specific content. Output failures arising from private intent omission are not model errors in the epistemic sense [7,8,26]: they are encoding failures on the user side that no downstream model improvement can address.

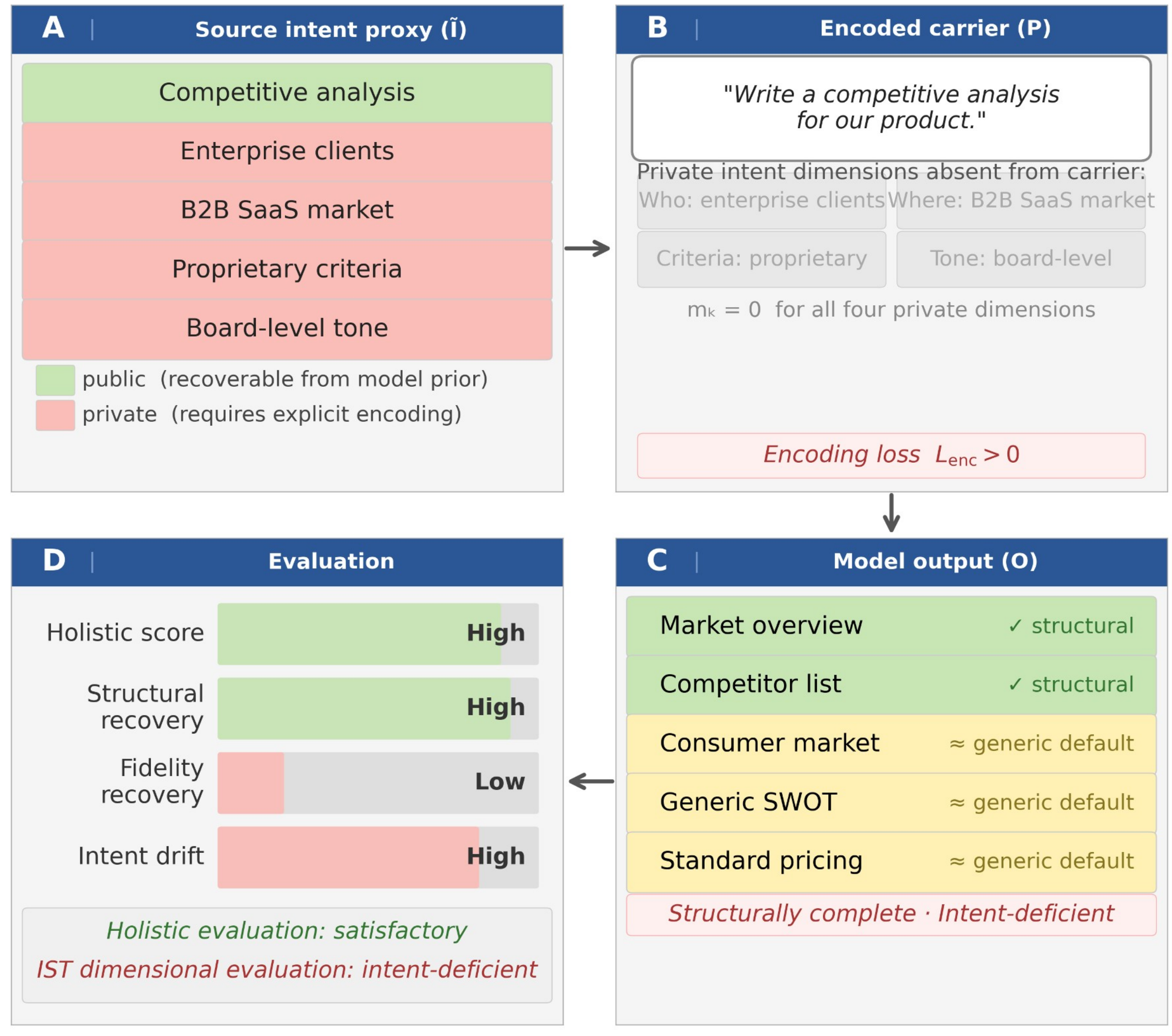

**Figure 2 |** Structural recovery without intent fidelity. A schematic example showing how an underspecified carrier can elicit a structurally complete but intent-deficient output. (A) The user's source intent proxy Ĩ includes five dimensions, four of which are private (enterprise audience, B2B market, proprietary criteria, board-level tone). (B)

The encoded carrier P is a minimal prompt that omits all four private dimensions (encoding mask $m_k = 0$). (C) The model output is structurally complete but fills private slots with generic defaults. (D) Holistic evaluation returns a high score because expected task structure is present, whereas IST dimensional evaluation detects low fidelity recovery and high intent drift — a structural–fidelity split that aggregate scoring conceals.

## 4. Empirical Grounding

The evidence base for IST is organised as a four-layer grounding chain, each layer instantiated by a companion empirical study publicly available on arXiv. These studies are used as grounding evidence rather than definitive validation: they establish empirical consistency with IST and motivate prospective tests, but do not by themselves exhaustively confirm the theory. Independent replication with alternative intent decompositions, annotators and task domains remains an important direction for future work. This grounding chain is also consistent with prior HCI and communication research showing that users systematically overestimate communicative success [22] and that conversational agents frequently fail to recover underspecified user expectations [23].

**Behavioural layer [19].** Structured intent encoding — using the 5W3H/PPS protocol as carrier — improves goal alignment across 540 outputs spanning three task domains, three Chinese LLMs, and three encoding conditions. Protocol-structured carriers produce consistently higher alignment scores than unstructured prompts, establishing the directional claim that encoding completeness — not prompt length or surface elaboration — drives the improvement. This is the IST prediction: higher explicit coverage of task-weighted dimensions reduces L_enc and improves recovery.

**Generalisability layer [20].** The alignment advantage of structured intent encoding survives translation across three languages (Chinese, English, Japanese) and six model families across 2,160 outputs. AI-expanded 5W3H specifications match human-authored specifications on goal alignment, with no significant language-by-encoding interaction. A weak-model compensation effect emerges: models with lower baseline capacity benefit more from structured encoding, consistent with IST's prediction that explicit encoding substitutes for limited prior-based recovery on dimensions that remain private under weaker priors.

**Protocol layer [21].** Structured intent behaves as a protocol-like communication mechanism across 3,240 outputs, outperforming alternative framework comparisons (CO-STAR, RISEN) on fidelity-sensitive dimensions and exhibiting a robustness pattern distinct from template formatting. The critical result is not that one framework outperforms others but that structured intent functions as a mechanism with measurable, stable properties across models — consistent with IST's protocol-layer account rather than a prompt-quality account.

**Measurement layer [12].** A systematic ablation study across 2,880 outputs and 2,520 dimension-model-language cells documents the structural-fidelity split, public–private asymmetry, human-validated metric dissociation, and weight-tolerance plateau; Figure 2 illustrates the split mechanism schematically. In Chinese, 25.7% of outputs fell in the split zone (GA = 5 while f-ICMw < 0.8); in English, 58.6% — a language difference consistent with training-density modulation of the public–private boundary. Across the ablation set, 31.5% of cells were classified as public-regime and 68.5% as private-regime, with the travel-planning domain showing the highest public rate and business/technical domains showing the highest private rate — consistent with IST's prediction that the public–private boundary tracks task conventionality. Human raters (n = 60 outputs) awarded split-zone outputs a mean GA of 3.12 versus LLM-judge scores of 5.0 for the same outputs; human–LLM agreement on the dimensional f-ICMw instrument was $\rho = 0.695$, compared with $\rho = 0.251$ on holistic GA — confirming that the structural-fidelity split is not a measurement artefact. Under severe weight inversion, fidelity degraded in 100% of domain-by-model cells; moderate misalignment produced scores within 0.02 of the matched baseline — the plateau-plus-cliff pattern IST predicts from its non-linear encoding sensitivity account.

Together, these four layers establish that structured intent encoding improves alignment (behavioural), the effect generalises across languages and models (generalisability), structured intent functions as a protocol-like mechanism rather than a template trick (protocol), and holistic evaluation systematically conceals the fidelity failures that dimensional instruments detect (measurement). IST provides the unified theoretical account of why all four phenomena co-occur. All underlying empirical materials are publicly available and can be inspected and reproduced independently, as described in the Data availability statement and Supplementary Note 1.

# 5. Implications and Research Agenda

## 5.1 From prompt engineering to intent-protocol design

The standard account of prompt engineering treats it as an optimisation problem over text: find the wording that elicits the best response [4,30,31]. IST offers a different framing. What varies across prompt styles is not primarily linguistic surface but encoding completeness: the degree to which the carrier P carries the user's source intent across its full dimensional profile. Good prompts minimise task-weighted encoding loss L_enc — particularly for high-weight dimensions in the private regime, where model-side recovery cannot substitute for explicit encoding. Protocol design should therefore prioritise private-dimension coverage, and evaluation of protocol effectiveness should use dimensional fidelity instruments rather than holistic quality scores. IST predicts that the advantage of structured protocols over unstructured prompting is concentrated on private dimensions and does not diminish as model capability increases; only the public-dimension advantage is partially substitutable by a richer model prior.

## 5.2 Evaluation beyond holistic alignment

A model can be fluent, helpful, and preferred by holistic evaluators while still failing to preserve user-specific intent. This is the systematic consequence of the structural-fidelity split, which the grounding evidence shows to be frequent enough to distort aggregate quality assessments across conditions, languages, and model tiers. Holistic alignment metrics — human rating scales, benchmark-level aggregate evaluations, and LLM-as-judge systems [27,28,29] — are insufficient as the sole evaluation instrument when the research question concerns intent preservation. Benchmark evaluation systems such as MT-Bench [28] and HELM [10] measure response quality against prompt surface or reference answers, but cannot distinguish a response that fills structural slots generically from one that preserves the user's private, individual-specific constraints; the public–private decomposition IST introduces is absent from these frameworks by design. IST motivates a complementary evaluation architecture with at minimum two layers: a structural layer measuring dimensional slot-filling and a fidelity layer measuring specific content reproduction. The structural-fidelity split makes both necessary, because they can diverge in ways that aggregate scores conceal. IST further predicts that alignment training on holistic preference signals [5,6] will not reliably converge toward private intent preservation: the training signal is insensitive to the structural-fidelity distinction that separates generic completion from user-specific intent fidelity.

## 5.3 Toward a computable, auditable intent layer

The deepest implication of IST is architectural. Current AI systems process prompts without representing what they carry. IST establishes the theoretical basis for an explicit intent layer that could record, for each interaction: which intent dimensions were explicitly encoded; which were absent from the carrier; which are estimated to be private and therefore at risk of fidelity loss; which were structurally recovered and which were fidelity-preserved; and what the resulting intent drift score is. Such a record makes human–AI interaction computable, evaluable, and — in high-stakes domains such as healthcare, education, law, and enterprise decision-making — auditable. The distinction IST draws between encoding failure and model error provides the vocabulary for attributing responsibility: a mismatch arising from absent private intent is not necessarily an epistemic hallucination, and addressing it requires protocol-side intervention, not further model training alone. Such an intent layer would make human–AI interaction not merely more effective, but governable. In this sense, IST is not a prompt-engineering theory but a protocol theory for governable human–AI interaction.

IST is related to, but distinct from, classical theories of communication and interaction. Gricean pragmatics [14] explains how speakers rely on cooperative inference when leaving information implicit, while Norman's gulfs of execution and evaluation [32] describe the distance between user goals, system actions and perceived outcomes. IST differs by making this gap computational and measurable in LLM interaction: it separates latent source intent, observable proxy, encoded carrier and generated output, and assigns dimension-level weights, encoding masks and fidelity scores to the transmission process. IST does not replace these frameworks; it operationalises their central insight for intent preservation in generative AI systems.

### 5.4 Falsifiable research agenda

IST yields five falsifiable predictions that extend beyond the grounding evidence. First, the structural-fidelity gap should be larger in task domains with higher private-dimension concentrations (business, personal) than in conventionalised domains (travel, standard formats), across model tiers. Second, capability improvements should produce larger fidelity gains on public dimensions than on private ones, producing a detectable capability-by-type interaction in factorial designs. Third, explicit encoding of private dimensions should narrow cross-model variance in fidelity recovery relative to omission. Fourth, the weight-tolerance plateau should be broader for predominantly public tasks and narrower for predominantly private tasks. Fifth, cross-modal analogues of the structural-fidelity split should appear in image and audio generation settings when the same ablation design is applied and public–private classifications are pre-registered. A minimal prospective test would pre-register public and private dimensions for a set of text-to-image tasks, ablate one dimension at a time from the prompt, and compare structural object recovery with fidelity to the omitted user-specific constraint.

Multi-intent composition, agentic tool use, long-term memory, and multimodal generation are natural extensions of IST, but they require separate formalisation and prospective validation beyond the single-turn scope established here.

## Data availability

This paper presents a theoretical framework and does not report a newly collected standalone experiment. Its empirical grounding draws on a publicly released evidence repository associated with the four text-generation grounding layers discussed in the manuscript. The repository contains prompt specifications, generated outputs, scoring files, human-evaluation materials, proxy annotation files, weight-perturbation data and analysis scripts supporting the behavioural, generalisability, protocol and measurement layers. The repository also includes additional exploratory multimodal materials, which are provided for transparency and future research continuity but are not used as grounding evidence for the claims made in this paper. The data are available at https://github.com/PGlarry/prompt-protocol-specification under CC BY 4.0. Additional materials are available from the corresponding author upon reasonable request.

## Code availability

Analysis and scoring scripts associated with the empirical grounding studies are available in the public repository at https://github.com/PGlarry/prompt-protocol-specification. These scripts support reproduction of the aggregate statistics and evaluation analyses summarised in the manuscript. No additional proprietary analysis code is required to interpret the results reported in this paper.

## Competing interests

G.P. is the creator of the 5W3H/PPS structured prompting framework and co-founder of Huizhou Lateni AI Technology Co., Ltd., which develops software tools related to structured prompt authoring. This potential competing interest has been disclosed. No other competing interests are declared.

## Funding

This research received no specific grant from any funding agency in the public, commercial, or not-for-profit sectors.

## Author contributions

G.P. conceived the Intent Signal Theory framework, developed the formal definitions and theorems, designed and conducted the companion empirical studies, and wrote the manuscript.

## Acknowledgements

The author thanks the open-source and research communities whose work on large language models, evaluation methodology, and information theory provided the foundation for this study.

## Correspondence

Correspondence and requests for materials should be addressed to Gang Peng (peng@hzu.edu.cn).

## Supplementary Note 1 | Public Empirical Evidence Repository

*Manuscript: Intent Signal Theory: A Computational Framework for Intent-State Control in Human-AI Interaction*

Repository: https://github.com/PGlarry/prompt-protocol-specification

## 1. Purpose of the Repository

This repository provides the empirical grounding materials that support the evidence chain discussed in the IST paper. It is released publicly to enable independent inspection, verification, and reproduction of the aggregate analyses underlying the grounding evidence. The repository does not replace journal peer review; it makes the underlying materials transparent and accessible so that the core claims of IST can be examined independently of the companion preprints cited in the manuscript.

## 2. Repository Structure and Evidence Mapping

The table below maps each grounding layer to the corresponding repository module. Paths are relative to dataset/data/ in the repository root.

| Grounding layer | Manuscript § | Repository module | Scale | Main variables | Role in present manuscript |
|---|---|---|---|---|---|
| Behavioural | §4, [19] | paper1/ | 540 outputs | GA, encoding condition | Grounding evidence |
| Generalisability | §4, [20] | paper2/ | 2,160 outputs | Language, model, GA | Grounding evidence |
| Protocol | §4, [21] | paper3/ | 3,240 outputs | Framework, model, GA | Grounding evidence |
| Measurement | §4, [12] | structural_fidelity_split/ | 2,880 + aux. | GA, f-ICMw, s-ICMw | Grounding evidence |
| Multimodal (expl.) | - | paper_image/ | - | Image tasks | *Exploratory; NOT used for present claims* |
| PPS-Bench | Background | PPS-Bench/ | 4,440+ records | Multi-lang. benchmark | Background; not central evidence |

## 3. Measurement-Layer Materials (Detailed)

### 3.1 Ablation study (01_ablation/)

**Design:** 30 tasks x 3 domains x 8 conditions (FULL + 7 single-dimension ablations) x models (6 for ZH, 3 each for EN and JA)

**Models:** DeepSeek-V3, Qwen-Max, Kimi, Claude Sonnet 4, GPT-4o, Gemini 2.5 Pro

**Format:** JSONL, one record per output: prompt, output text, GA, f-ICMw, s-ICMw, condition label, dimension-mask vector

**Total:** 2,880 outputs across 2,520 dimension-model-language cells

### 3.2 Human evaluation (02_human_evaluation/)

**Design:** Stratified sample of N = 60 outputs; two independent human raters (Rater A: 5W3H-familiar; Rater B: naive to framework)

**Materials:** Anonymised annotation sheets, answer key, analysis report, inter-rater reliability statistics

**Key finding:** Human raters awarded split-zone outputs a mean GA of 3.12 vs LLM-judge GA of 5.0; human-LLM agreement on f-ICMw was rho = 0.695 vs rho = 0.251 on holistic GA

### 3.3 Proxy annotation (03_proxy_annotation/)

**Design:** 210 task x dimension units; Claude Sonnet 4 and GPT-4o independently labelled each unit for public inferability; merged consensus labels used for public-private classification

**Materials:** Per-model annotation files, merged consensus file, classification report

### 3.4 Weight-perturbation experiment (04_weight_experiment/)

**Design:** v2: 240 outputs across weight conditions; v3_clean: 120 outputs with dual-judge verification (DeepSeek-V3, Qwen-Max, Kimi)

**Key finding:** Severe weight inversion degraded WAS in 100% of domain x model cells; moderate misalignment produced scores within 0.02 of the matched baseline (plateau-plus-cliff pattern)

### 3.5 Analysis scripts (analysis_scripts/)

Scripts for aggregating GA and f-ICMw scores, computing split-zone proportions, running public-private asymmetry analyses, and producing the figures reported in the manuscript. Entry point: README.md in the analysis_scripts/ directory.

## 4. Reproducibility Statement

All materials listed above are publicly available and can be downloaded, executed, and verified independently. Task definitions used to generate outputs are in tasks/tasks.json. Scoring files are organised by model and language under scores/. No proprietary software is required to reproduce the aggregate statistics reported in the manuscript.

These materials are publicly released for transparency and independent verification. They have not yet all completed journal peer review, and independent replication with alternative intent decompositions, annotators, and task domains remains an important direction for future work.

## 5. Evidence Status and Scope Limitations

**Evidence status.** The materials in this repository are pre-peer-review unless explicitly noted. They constitute public grounding evidence supporting the IST framework, not independently replicated findings.

**Scope.** This paper focuses on single-turn text-generation interactions. Extension to multi-turn, agentic, and multimodal settings is outside the current scope and is noted in Section 5.4 as a direction for future formalisation.

**Independence.** All data were generated by the corresponding author's research programme. Independent replication by external teams is an explicit goal of the falsifiable research agenda stated in Section 5.4.

## Supplementary Table 1 | Mapping between IST claims and public evidence modules

| Manuscript claim | Layer | Repository module | Scale | Key variables | Status |
|---|---|---|---|---|---|
| Structured encoding improves | Behavioural | paper1/ | 540 | GA, condition | arXiv + public data |

| Manuscript claim | Layer | Repository module | Scale | Key variables | Status |
|---|---|---|---|---|---|
| alignment | | | | | |
| Advantage generalises cross-language and cross-model | Generalisability | paper2/ | 2,160 | Language, model, GA | arXiv + public data |
| Structured intent functions as protocol, not template trick | Protocol | paper3/ | 3,240 | Framework, model, GA | arXiv + public data |
| Structural-fidelity split is systematic | Measurement | sf_split/ 01_ablation/ | 2,880 | GA, f-ICMw, s-ICMw | arXiv + public data |
| Split is not a measurement artefact (human validation) | Measurement | sf_split/ 02_human_eval / | N=60 | Human GA, LLM GA, f-ICMw | Public data |
| Public-private asymmetry via proxy annotation | Measurement | sf_split/ 03_proxy/ | 210 units | Inferability labels | Public data |
| Weight-tolerance plateau-plus-cliff | Measurement | sf_split/ 04_weight/ | 360 | WAS, weight condition | Public data |

Note: sf_split/ is shorthand for structural_fidelity_split/. All module paths are relative to dataset/data/ in the repository root.